\documentclass[english,aps,preprint,superscriptaddress]{revtex4-2}
\usepackage{lmodern}
\usepackage{lmodern}
\usepackage[T1]{fontenc}
\usepackage[utf8]{inputenc}
\usepackage{amsmath}
\usepackage{amsthm}
\usepackage{amssymb}
\usepackage{graphicx}
\usepackage{esint}

\makeatletter

\usepackage{amsthm}\usepackage{latexsym}\usepackage{bm}\usepackage{amsfonts}

\usepackage{babel}
\usepackage{enumitem}

\usepackage{hyperref}
\hypersetup{
	breaklinks = true,
    colorlinks = true,
    citecolor = {blue},
	urlcolor = {blue},
	linkcolor = {blue}
}

\makeatother

\usepackage{babel}
\begin{document}
\title{Geometric Electrostatic Particle-In-Cell Algorithm on Unstructured
Meshes}
\author{Zhenyu Wang, Hong Qin, Benjamin Sturdevant, Choong-Seock Chang}
\affiliation{Princeton Plasma Physics Laboratory, Princeton University, Princeton,
NJ 08543}
\begin{abstract}
We present a geometric Particle-in-Cell (PIC) algorithm on two-dimensional
(2D) unstructured meshes for studying electrostatic perturbations
in magnetized plasmas. In this method, ions are treated as fully kinetic
particles, and electrons are described by the adiabatic response.
The PIC method is derived from a discrete variational principle on
unstructured meshes. To preserve the geometric structure of the system,
the discrete variational principle requires that the electric field
is interpolated using Whitney 1-forms, the charge is deposited using
Whitney 0-forms, and the electric field is computed by discrete exterior
calculus. The algorithm has been applied to study the Ion Bernstein
Wave (IBW) in 2D magnetized plasmas. The simulated dispersion relations
of the IBW in a rectangular region agree well with theoretical results.
In a 2D circular region with the fixed boundary condition, the spectrum
and eigenmode structures of the IBW are determined from simulation.
We compare the energy conservation property of the geometric PIC algorithm
derived from the discrete variational principle with that of previous
PIC methods on unstructured meshes. The comparison shows that the
new PIC algorithm significantly improves the energy conservation property. 
\end{abstract}
\maketitle

\section{Introduction \label{sec:Introduction}}

Particle-In-Cell (PIC) simulation is an important tool for plasma
physics \citep{DAWSON1976,Dawson1983,Hockney1981,Potter1973,Birdsall1991}.
Structured mesh is easy to implement and widely used in PIC simulations.
On the other hand, some studies require modeling plasma behaviors
in specific and complex geometries, where unstructured meshes have
an unique advantage. Electrostatic (ES) PIC schemes on unstructured
meshes have been proposed \citep{Celik2003,Spirkin2004,Gatsonis2009,Day2011,Han2016}.
In these schemes, the shape function for interpolating the electric
field at particles' positions is identical to that for depositing
particles' charge to the grid points of unstructured meshes. Numerical
studies \citep{Langdon1970} showed that using the same shape function
for charge deposition and field interpolation restricts the grid size
to the Debye length. It is difficult to carry out large-scale simulations
for collisionless plasma using these ES PIC schemes. In previous PIC
methods, the charge-deposition algorithm and the field-interpolation
algorithm are independent. The shape functions for these two algorithms
can be chosen to be the same or different. There is no fundamental
guiding principle on how to design these two algorithms. In the present
study, we develop a geometric algorithm for ES PIC simulations with
adiabatic electrons on an unstructured mesh. Instead of select a shape
function based on intuition or experience, we derive the charge-deposition
and field-interpolation algorithm from an underpinning discrete variational
principle. 

Squire et al. \citep{Squire2012,Squire4748} first employed the methodology
of discrete variational principle \citep{lee82,Veselov1988,Marsden2001,qin2008variational,qin2009variational,Qin2020ML}
to derive an electromagnetic (EM) PIC algorithm on an unstructured
mesh. In this work, the technique of Whitney forms \citep{Whitney1957}
were introduced for the first time to deposit charge and current and
to interpolate fields. Discrete Exterior Calculus (DEC) \citep{Bossavit1998,Hirani2003}
was also applied to compute the electromagnetic field on the unstructured
mesh. It was demonstrated that the discrete variational principle
admits the discrete electromagnetic gauge symmetry and thus ensures
the discrete charge conservation \citep{Squire2012,Squire4748,Xiao2015,Xiao2018structure,Glasser_2020}.
Xiao et al. \citep{Xiao2015} developed an explicit high-order non-canonical
symplectic electromagnetic PIC schemes starting from the discrete
variational principles on a cubic mesh. High-order Whitney forms for
cubic meshes were constructed and used for the current deposition
and electromagnetic field interpolation. It was found that the ``shape
functions'' for current deposition and field interpolation are different,
and even for different components of the field the interpolation schemes
are different. Similar and subsequent studies \citep{xiao2013variational,xiao2015variational,xiao2017local,Xiao2019,Xiao2020,Zheng2020}
have also illustrated that discrete variational principles and Whitney
forms are useful tools in designing structure-preserving geometric
PIC algorithms \citep{Squire2012,Squire4748,Xiao2015,he2015hamiltonian,Qin2016,he2016hamiltonian,kraus2017gempic,burby2017finite,Morrison2017,Xiao2018structure,Li2019,Perse2020,Kormann2021}
. Even for partially-geometric PIC algorithms, the application of
Whitney forms has been found to be beneficial \citep{Moon2015}. 

In plasma physics, many reduced models are used. It is desirable to
apply structure-preserving geometric algorithms to these models as
well. One widely adopted reduced model is the Vlasov-Poisson system
with kinetic ions and adiabatic electrons. A structure-preserving
geometric PIC algorithm on a cubic mesh for this system was developed
recently \citep{Xiao2019}. The construction of the algorithm starts
from a field theory, i.e., a variational principle. As in the geometric
PIC algorithms for the Vlasov-Maxwell system, Whitney forms and variational
symplectic integrators are employed. In particular, the charge-deposition
and field-interpolation schemes were derived from the variational
principle for the electrostatic dynamics in the cubic mesh. 

We take a step-by-step approach to extend the structure-preserving
geometric PIC algorithm for the Vlasov-Poisson system reported in
Ref.\,\citep{Xiao2019} to unstructured meshes. In the present work,
we focus on the charge-deposition and field-interpolation methods
and have not implemented the symplectic integrator. The Lagrangian
of the system is discretized on an unstructured mesh, and the charge-deposition
and field-interpolation methods are derived from the discrete variational
principle on the unstructured mesh using Whitney forms. In place of
a more structure-preserving symplectic integrator \citep{he2015hamiltonian,he2016hamiltonian,He2017},
the Boris algorithm \citep{Boris1970}, which preserves the phase
space volume \citep{Qin2013,he2015volume,zhang2015volume,he16-172,He2016HigherRela},
is adopted to push particles. Our purpose here is to demonstrate how
to design effective charge-deposition and field-interpolation algorithms
on an unstructured mesh using the discrete field theory and Whitney
forms. To validate the new PIC algorithm, we compare the dispersion
relation of the Ion Bernstein Wave (IBW) from the PIC simulation with
the theory \citep{Sturdevant2016,Sturdevant2017} in a two-dimensional
(2D) periodic plasma. Eigenmode structures of the IBW in a 2D circular
geometry with Dirichlet boundary conditions are also simulated. We
compare the simulation results with those of the conventional methods
\citep{Celik2003,Spirkin2004} on the same unstructured mesh, and
find that our method is able to significantly reduce the energy error
of the simulations.

The paper is organized as follows. In Sec. II, the geometric electrostatic
PIC algorithm with fully kinetic ions and adiabatic electrons on an
unstructured mesh is derived. Simulations of the IBW in an infinite
2D geometry and a 2D circular geometry are presented in Sec.\,III.
We compare the energy conservation property of our algorithm with
that of previous methods in Sec.\,IV. 

\section{Geometric electrostatic PIC algorithm on an unstructured mesh\label{sec:Model}}

In this section, we extend the geometric electrostatic PIC algorithm
reported in Ref.\,\citep{Xiao2019} to an unstructured mesh. The
model treats ions as fully kinetic 6D particles. The response of the
electrons is adiabatic \citep{Horton1999,Weiland2012,Sturdevant2016,hu2018fully,miecnikowski2018nonlinear},
which, with the quasi-neutrality condition, leads to 
\begin{equation}
-\frac{q_{i}}{q_{e}}n_{i}=n_{e0}\textrm{exp}(-\frac{q_{e}\phi}{T_{e}}),\label{eq:adiabatic}
\end{equation}
where $n_{i}$ is the ion density, $n_{e0}$ the background electron
density, $\phi$ the electric potential, $T_{e}$ the electron temperature,
and $q_{e}$ and $q_{i}$ are electron and ion charges respectively.
The action integral of the system is \citep{Xiao2019} 
\begin{align}
S(\mathbf{x},\phi) & =\int dtL(\mathbf{x},\phi),\\
L & =\int d\mathbf{x}d\mathbf{v}f_{i}(\mathbf{x},\mathbf{v})\left[\frac{1}{2}m_{i}\dot{\mathbf{x}}^{2}+q_{i}\dot{\mathbf{x}}\cdot\mathbf{A}_{0}(\mathbf{x},t)\right]\nonumber \\
 & -q_{i}\left[\phi(\mathbf{x},t)+\phi_{0}(\mathbf{x},t)\right]+n_{e0}T_{e}\textrm{exp}\left[-\frac{q_{e}\phi(\mathbf{x},t)}{T_{e}}\right],
\end{align}
where $\mathbf{x}=\mathbf{x}(\mathbf{x}_{0},\mathbf{v}_{0},t)$, $f_{i}$
is the ion distribution function, $m_{i}$ is ion mass, and $\mathbf{A}_{0}$
and $\phi_{0}$ are the external vector and scalar potentials. The
dynamics of the system is governed by the Euler-Lagrange equations,
\begin{align}
\frac{\delta S}{\delta\phi} & =0,\label{eq:ELphi}\\
\frac{\delta S}{\delta\mathbf{x}} & =0.\label{eq:ELx}
\end{align}
Equation (\ref{eq:ELphi}) links the electric potential $\phi$ and
the ion charge density $\rho_{i}$, 
\begin{equation}
\phi=-\frac{T_{e}}{q_{e}}\textrm{log}(-\frac{\rho_{i}}{q_{e}n_{e0}}),\label{eq:Phi}
\end{equation}
where $\rho_{i}={\displaystyle \int d\mathbf{v}q_{i}f_{i}(\mathbf{x},\mathbf{v})}$.
Equation (\ref{eq:Phi}) recovers the electron adiabatic response
and charge-neutrality condition in Eq.\,(\ref{eq:adiabatic}). Equation
(\ref{eq:ELx}) gives equation of motion for particles, 
\begin{equation}
\ddot{\mathbf{x}}=\frac{q_{i}}{m_{i}}\left[\mathbf{E}_{0}(\mathbf{x},t)+\mathbf{E}(\mathbf{x},t)+\dot{\mathbf{x}}\times\mathbf{B}_{0}(\mathbf{x},t)\right],
\end{equation}
where $\mathbf{E}_{0}(\mathbf{x},t)$ and $\mathbf{B}_{0}(\mathbf{x},t)$
are the external electromagnetic field, and $\mathbf{E}(\mathbf{x},t)=-\nabla\phi(\mathbf{x},t)$
is the perturbed electrostatic field.

We use the method introduced in Refs.\,\citep{Squire2012,Xiao2019}
to discretize the action integral by particles and Whitney forms \citep{Whitney1957}
on a 2D unstructured triangular mesh. The discrete action integral
$S_{d}$ can be written as 
\begin{align}
S_{d}(\mathbf{x}_{p},\phi_{I}) & =\int L_{d}(\mathbf{x}_{p},\phi_{I})dt,\\
L_{d}(\mathbf{x}_{p},\phi_{I}) & =\sum_{p}\Big[\frac{1}{2}m_{i}\dot{\mathbf{x}}_{p}^{2}+q_{i}\dot{\mathbf{x}}_{p}\mathbf{A}_{0}(\mathbf{x}_{p})-q_{i}\sum_{I}W_{\sigma_{0},I}(\mathbf{x}_{p})\phi_{I}\nonumber \\
 & -q_{i}\phi_{0}(\mathbf{x}_{p})\Big]+\sum_{I}n_{e0,I}T_{e,I}\textrm{exp}(-\frac{q_{e}\phi_{I}}{T_{eI}}),
\end{align}
where $I$ is triangular vertex index, $\mathbf{x}_{p}$ the particle
position of the $p$-th particle, $\phi_{I}$ is the electric potential
defined on the triangular vertex, and $W_{\sigma_{0},I}$ is the Whitney
0-form interpolating the value of $\phi$ in the continuous space
using $\phi_{I}$. The discrete action integral is the same as in
Ref.\,\citep{Xiao2019}, except that here it is on a 2D unstructured
triangular mesh.

Variations of $S_{d}$ with respect to $\phi_{I}$ and $\mathbf{x}_{p}$
lead to the equations of motion of the electrostatic system, 
\begin{equation}
\frac{\delta S}{\delta\phi_{I}}=0,\label{eq:dELphi}
\end{equation}
\begin{equation}
\frac{\delta S}{\delta\mathbf{x}_{p}}=0.\label{eq:dELx}
\end{equation}
Equation (\ref{eq:dELphi}) gives
\begin{equation}
\phi_{I}=-\frac{T_{e,I}}{q_{e}}\textrm{log}(-\frac{\rho_{I}}{q_{e}n_{e0,I}}),\label{eq:disphi}
\end{equation}
where
\begin{equation}
\rho_{I}=q_{i}W_{\sigma_{0},I}(\mathbf{x}_{p})\label{eq:rhoi}
\end{equation}
is the charge density on the triangular vertex. Equation (\ref{eq:dELx})
is the governing equation for ion dynamics, 
\begin{equation}
\ddot{\mathbf{x}}_{p}=\frac{q_{i}}{m_{i}}\Big[\mathbf{E}_{0}(\mathbf{x}_{p},t)+\dot{\mathbf{x}}_{p}\times\mathbf{B}_{0}(\mathbf{x}_{p},t)-\frac{\partial}{\partial\mathbf{x}_{p}}\sum_{I}W_{\sigma_{0},I}(\mathbf{x}_{p})\phi_{I}\Big].\label{eq:dpm}
\end{equation}
The last term of Eq.\,(\ref{eq:dpm}) is the derivative of $W_{\sigma_{0},I}$
with respect to $\mathbf{x}_{p}$ in the continuous space. According
to the property of Whitney forms \citep{Whitney1957,Hirani2003,Squire2012,Xiao2015},
\begin{equation}
\nabla\sum_{I}W_{\sigma_{0},I}(\mathbf{x}_{p})\phi_{I}=\sum_{J}W_{\sigma_{1},J}(\mathbf{x}_{p})\sum_{I}\nabla_{dJ,I}\phi_{I}.\label{eq:dW}
\end{equation}
where $\sum_{I}\nabla_{dJ,I}\phi_{I}$ is the discrete gradient of
$\phi_{I}$, and $W_{\sigma_{1},J}$ is the Whitney 1-form that interpolates
a continuous 1-form from the discrete 1-form defined on the triangular
edge. The construction of $W_{\sigma_{0},I}$, $W_{\sigma_{1},J}$
and $\nabla_{dJ,I}$ will be discussed shortly after. With property
(\ref{eq:dW}), Eq.\,(\ref{eq:dpm}) becomes
\begin{equation}
\ddot{\mathbf{x}}_{p}=\frac{q_{i}}{m_{i}}\Big[\mathbf{E}_{0}(\mathbf{x}_{p},t)+\dot{\mathbf{x}}_{p}\times\mathbf{B}_{0}(\mathbf{x}_{p},t)+\sum_{J}W_{\sigma_{1},J}(\mathbf{x}_{p})\mathbf{E}_{J}\Big],\label{eq:dpmW}
\end{equation}
and 
\begin{equation}
\mathbf{E}_{J}=-\sum_{I}\nabla_{dJ,I}\phi_{I},\label{eq:elec_edge}
\end{equation}
where $\mathbf{E}_{J}$ is the discrete electrical field defined on
the triangular edge labeled by $J$. We want to emphasize again that,
similar to the scenario in a cubic mesh \citep{Xiao2019}, without
Whitney forms and DEC, it is difficult to calculate on the electric
field on an unstructured mesh to advance particles' positions. 

As is well-known, the key parts of a PIC method include charge deposition,
solving discrete field, and field interpolation, which are encapsulated
in a systematic way in Eqs.\,(\ref{eq:rhoi}), (\ref{eq:elec_edge}),
and (\ref{eq:dpmW}), respectively. Once $W_{\sigma_{0},I}$, $W_{\sigma_{1},J}$
and $\nabla_{dJ,I}$ are chosen, the PIC algorithm is defined. 

\begin{figure}
\includegraphics[scale=0.5]{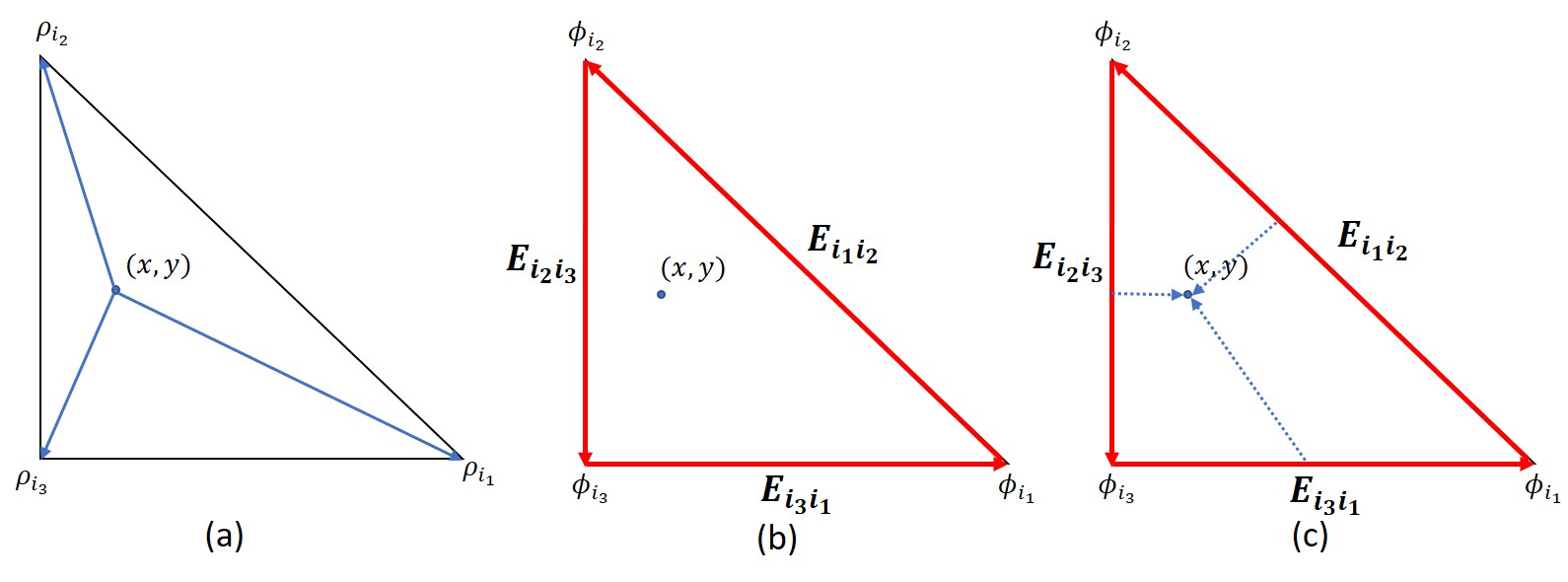}

\caption{\label{fig:dfi}The PIC algorithm on a triangle. (a) Depositing a
particle's charge to the triangular vertices using Whitney 0-forms
$W_{\sigma_{0},I}$. (b) Computing $\mathbf{E}_{J}$ on the edge with
the discrete gradient operator $\nabla_{dJ,I}$. (c) Interpolating
$\mathbf{E}_{J}$ form edges to the particle's location through Whitney
1-forms $W_{\sigma_{1},J}$.}
\end{figure}

\begin{figure}
\includegraphics[scale=0.4]{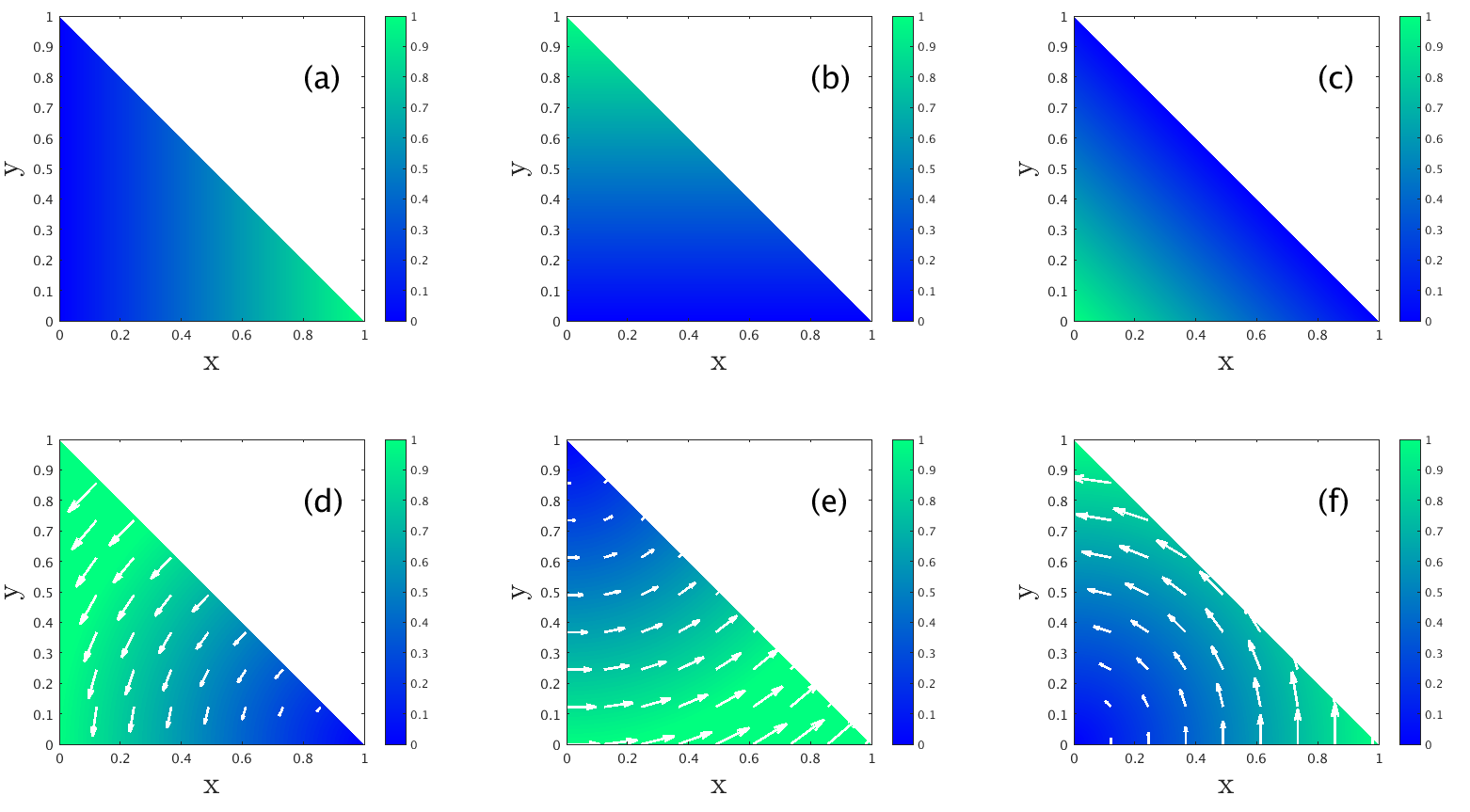}

\caption{\label{fig:WF}The values of the Whitney 0-forms and 1-forms. The
density plot of the value of (a) $W_{\sigma_{0},i_{1}}(x,y)$, (b)
$W_{\sigma_{0},i_{2}}(x,y)$, and (c) $W_{\sigma_{0},i_{3}}(x,y)$.
The amplitude density plot and the quiver plot of (d) $W_{\sigma_{1},i_{2}i_{3}}(x,y)$,
(e) $W_{\sigma_{1},i_{3}i_{1}}(x,y)$, and (f) $W_{\sigma_{1},i_{1}i_{2}}(x,y)$.}
\end{figure}

Now we describe in details the construction of $W_{\sigma_{0},I}$,
$W_{\sigma_{1},J}$ and $\nabla_{dJ,I}$ on the triangular mesh. Figure
\ref{fig:dfi}(a) shows a particle in a triangle, $(x,y)$ is the
position of the particle $p$, and the three vertices of the $i$-th
triangle, $i_{1}$, $i_{2}$, $i_{3}$, have coordinates $(x_{i_{1}},y_{i_{1}})$,
$(x_{i_{2}},y_{i_{2}})$, $(x_{i_{3}},y_{i_{3}})$, respectively.
To deposit charge at each vertex according to Eq.\,(\ref{eq:rhoi}),
we need to specify the Whitney 0-forms, which are chosen to be linear
barycentric functions. For $(x,y)$ inside the triangle, the Whitney
0-forms are

\begin{align}
W_{\sigma_{0},i_{1}}(x,y) & =\frac{(y_{i_{2}}-y_{i_{3}})(x-x_{i_{3}})+(x_{i_{3}}-x_{i_{2}})(y-y_{i_{3}})}{(x_{i_{1}}-x_{i_{3}})(y_{i_{2}}-y_{i_{3}})+(x_{i_{3}}-x_{i_{2}})(y_{i_{1}}-y_{i_{3}})},\label{eq:W01}\\
W_{\sigma_{0},i_{2}}(x,y) & =\frac{(y_{i_{3}}-y_{i_{1}})(x-x_{i_{3}})+(x_{i_{1}}-x_{i_{3}})(y-y_{i_{3}})}{(x_{i_{1}}-x_{i_{3}})(y_{i_{2}}-y_{i_{3}})+(x_{i_{3}}-x_{i_{2}})(y_{i_{1}}-y_{i_{3}})},\label{eq:W02}\\
W_{\sigma_{0},i_{3}}(x,y) & =1-W_{\sigma_{0},i_{1}}(x,y)-W_{\sigma_{0},i_{2}}(x,y).\label{eq:W03}
\end{align}
When $(x,y)$ is outside the triangle, all the Whitney 0-forms vanishes.
Note that $W_{\sigma_{0},i_{1}}(x,y)$, $W_{\sigma_{0},i_{2}}(x,y)$
and $W_{\sigma_{0},i_{3}}(x,y)$ are the areas of the triangles $\Delta pi_{2}i_{3}$,
$\Delta pi_{3}i_{1}$, and $\Delta pi_{1}i_{2}$, respectively. Thus,
the weighting of charge deposition with respect to a vertex is the
area weighting of the triangle enclosed by the particle and the opposite
edge of the vertex. The density plots of $W_{\sigma_{0},i_{1}}(x,y)$,
$W_{\sigma_{0},i_{2}}(x,y)$ and $W_{\sigma_{0},i_{3}}(x,y)$ are
shown in Fig.\,\ref{fig:WF}(a)-(c). Figure \ref{fig:WF}(a) shows
$W_{\sigma_{0},i_{1}}(x,y)$ approaching to 1 as the particle is close
to the vertex $i_{1}$. The chosen Whitney 0-forms in Eqs.\,(\ref{eq:W01})-(\ref{eq:W03})
obviously satisfy the condition, 
\begin{align}
{\displaystyle \sum_{I=i_{1},i_{2},i_{3}}W_{\sigma_{0},I}(\mathbf{x}_{p})} & =1.
\end{align}
In the 2D triangular mesh, 1-forms, such as $\mathbf{E}_{J}$, are
defined on the triangular edges. And the index $J$ for the edges
consists of an ordered pair of indices of the vertices. For example,
$J=i_{1}i_{2}$ labels the oriented edge from $i_{1}$ to $i_{2}$,
as shown in Figure \ref{fig:dfi}(b). The discrete gradient operator
$\nabla_{dJ,I}$ consistent with Eq.\,(\ref{eq:dW}) is 
\begin{align}
\nabla_{di_{1}i_{2},I} & =\delta_{i_{2}I}-\delta_{i_{1}I}.
\end{align}

According to the definition of Whitney forms \citep{Whitney1957},
the Whitney 1-form on unstructured triangular mesh is 

\begin{equation}
W_{\sigma_{1},j^{\prime}j}=W_{\sigma_{0},j}\nabla W_{\sigma_{0},j^{\prime}}-W_{\sigma_{0},j^{\prime}}\nabla W_{\sigma_{0},j}.
\end{equation}

For the triangular mesh, the expressions of $W_{\sigma_{1},j^{\prime}j}(x,y)$
for $j^{\prime},j=i_{1},i_{2},i_{3}$ and $j^{\prime}\neq j$ are
\begin{align}
W_{\sigma_{1},i_{1}i_{2}}(x,y) & =\Bigg[\frac{(y_{i_{3}}-y_{i_{1}})W_{\sigma_{0},i_{1}}(x,y)-(y_{i_{2}}-y_{i_{3}})W_{\sigma_{0},i_{2}}(x,y)}{(x_{i_{1}}-x_{i_{3}})(y_{i_{2}}-y_{i_{3}})+(x_{i_{3}}-x_{i_{2}})(y_{i_{1}}-y_{i_{3}})},\nonumber \\
 & \frac{(x_{i_{1}}-x_{i_{3}})W_{\sigma_{0},i_{1}}(x,y)-(x_{i_{3}}-x_{i_{2}})W_{\sigma_{0},i_{2}}(x,y)}{(x_{i_{1}}-x_{i_{3}})(y_{i_{2}}-y_{i_{3}})+(x_{i_{3}}-x_{i_{2}})(y_{i_{1}}-y_{i_{3}})}\Bigg],\label{W12}\\
W_{\sigma_{1},i_{2}i_{3}}(x,y) & =\Bigg[\frac{(y_{i_{1}}-y_{i_{2}})W_{\sigma_{0},i_{2}}(x,y)-(y_{i_{3}}-y_{i_{1}})W_{\sigma_{0},i_{3}}(x,y)}{(x_{i_{1}}-x_{i_{3}})(y_{i_{2}}-y_{i_{3}})+(x_{i_{3}}-x_{i_{2}})(y_{i_{1}}-y_{i_{3}})},\nonumber \\
 & \frac{(x_{i_{2}}-x_{i_{1}})W_{\sigma_{0},i_{2}}(x,y)-(x_{i_{1}}-x_{i_{3}})W_{\sigma_{0},i_{3}}(x,y)}{(x_{i_{1}}-x_{i_{3}})(y_{i_{2}}-y_{i_{3}})+(x_{i_{3}}-x_{i_{2}})(y_{i_{1}}-y_{i_{3}})}\Bigg],\label{W23}\\
W_{\sigma_{1},i_{3}i_{1}}(x,y) & =\Bigg[\frac{(y_{i_{2}}-y_{i_{3}})W_{\sigma_{0},i_{3}}(x,y)-(y_{i_{1}}-y_{i_{2}})W_{\sigma_{0},i_{1}}(x,y)}{(x_{i_{1}}-x_{i_{3}})(y_{i_{2}}-y_{i_{3}})+(x_{i_{3}}-x_{i_{2}})(y_{i_{1}}-y_{i_{3}})},\nonumber \\
 & \frac{(x_{i_{3}}-x_{i_{2}})W_{\sigma_{0},i_{3}}(x,y)-(x_{i_{2}}-x_{i_{1}})W_{\sigma_{0},i_{1}}(x,y)}{(x_{i_{1}}-x_{i_{3}})(y_{i_{2}}-y_{i_{3}})+(x_{i_{3}}-x_{i_{2}})(y_{i_{1}}-y_{i_{3}})}\Bigg],\label{W31}
\end{align}
for $(x,y)$ inside the triangle. All $W_{\sigma_{1},j^{\prime}j}(x,y)$
vanishes when $(x,y)$ is outside the triangle. The amplitude density
plot and the quiver plot of $W_{\sigma_{1},i_{2}i_{3}}(x,y)$ are
showed in Fig.\,\ref{fig:WF}(d), where the amplitude approaches
zero when a particle is close to the opposite vertex of the edge and
maximizes when the particle approaches the edge. Figures \ref{fig:WF}(e)
and (f) plot the value and direction of $W_{\sigma_{1},i_{3}i_{1}}(x,y)$
and $W_{\sigma_{1},i_{1}i_{2}}(x,y)$.

After $W_{\sigma_{0},I}$, $W_{\sigma_{1},J}$ and $\nabla_{dJ,I}$
are chosen, the discrete electric potential $\phi_{I}$ at each vertex
can be calculated by Eqs.\,(\ref{eq:rhoi}) and (\ref{eq:disphi}),
and the electric field on the edges according to Eq.\,(\ref{eq:elec_edge})
is 
\begin{equation}
E_{i_{1}i_{2}}=\phi_{i_{2}}-\phi_{i_{1}},
\end{equation}
where $E_{i_{1}i_{2}}$ is a discrete 1-form, we denote it by a boldface
symbol in Fig.\,\ref{fig:dfi} to following the convention of physicists.
Particles' positions and velocities are advanced according to Eq.\,(\ref{eq:dpmW}),
which interpolates the electrical field at $\mathbf{x}_{p}$ as $\textbf{\emph{E}}(\mathbf{x}_{p})=\sum_{J}W_{\sigma_{1},J}(\mathbf{x}_{p})E_{J}$
using Whitney 1-forms. Note that by definition $W_{\sigma_{1},J}(\mathbf{x}_{p})$
is a vector and $E_{J}$ is a scalar. This process is illustrated
in Fig.\,\ref{fig:dfi}(c). In the current implementation, Eq.\,(\ref{eq:dpmW})
is integrated by the Boris algorithm \citep{Boris1970}, which preserves
the phase space volume \citep{Qin2013,he2015volume,he16-172,zhang2015volume,ellison2015comment,He2016HigherRela}
although is not symplectic. 

\section{Ion Bernstein waves on 2D unstructured meshes \label{sec:Examples}}

The PIC method described in Sec.\ref{sec:Model} is used to examine
Ion Bernstein wave (IBW) \citep{Bernstein1958} in 2D uniform plasmas.
The simulations are carried out in a rectangular region with periodic
boundary conditions and in a 2D circular region with fixed boundary
conditions. 

\subsection{IBW with periodic boundary conditions in a rectangular region \label{subsec:IBW1}}

\begin{figure}
\includegraphics[scale=0.7]{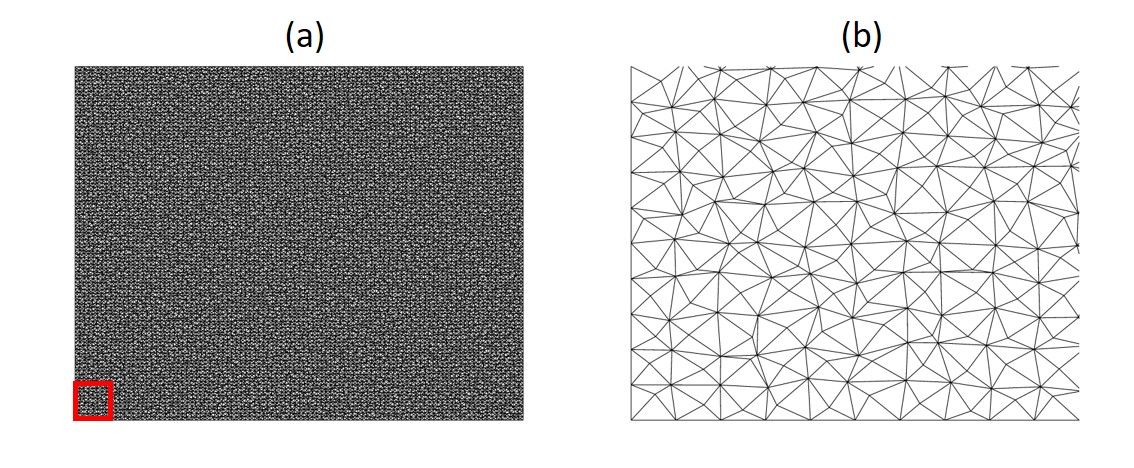}

\caption{\label{fig:mrec}(a) The rectangular simulation domain. (b) The zoom-in
of the red rectangular region in (a).}
\end{figure}

First, the IBW in a rectangular region of an uniform plasma is examined
on an unstructured mesh with periodic boundary conditions. The simulation
domain is shown in Fig.\,\ref{fig:mrec}(a), and Fig.\,\ref{fig:mrec}(b)
is the zoom-in of the red region in Fig.\,\ref{fig:mrec}(a). The
simulation domain has 20201 vertices and 40000 triangles. To implement
the periodic boundary conditions, each vertex at the left boundary
is identified with the corresponding vertex at the right boundary.
Similar identification is imposed for the top and bottom boundaries.
The ion mass is $m_{i}$=1.67$\times$10$^{-27}$kg and charge is
$q_{i}$=1.6$\times$10$^{-19}$C, the initial ion density is $n_{i0}$=10$^{20}$/$m^{3}$,
the electron and ion temperatures are 1000eV, and the out-of-plane
background magnetic field is $B_{0}$=2T. The simulation timestep
is $\Delta t$=0.01$/\Omega_{i}$, and the total number of simulation
particles is 5.12$\times$10$^{7}$.

\begin{figure}
\includegraphics[scale=0.6]{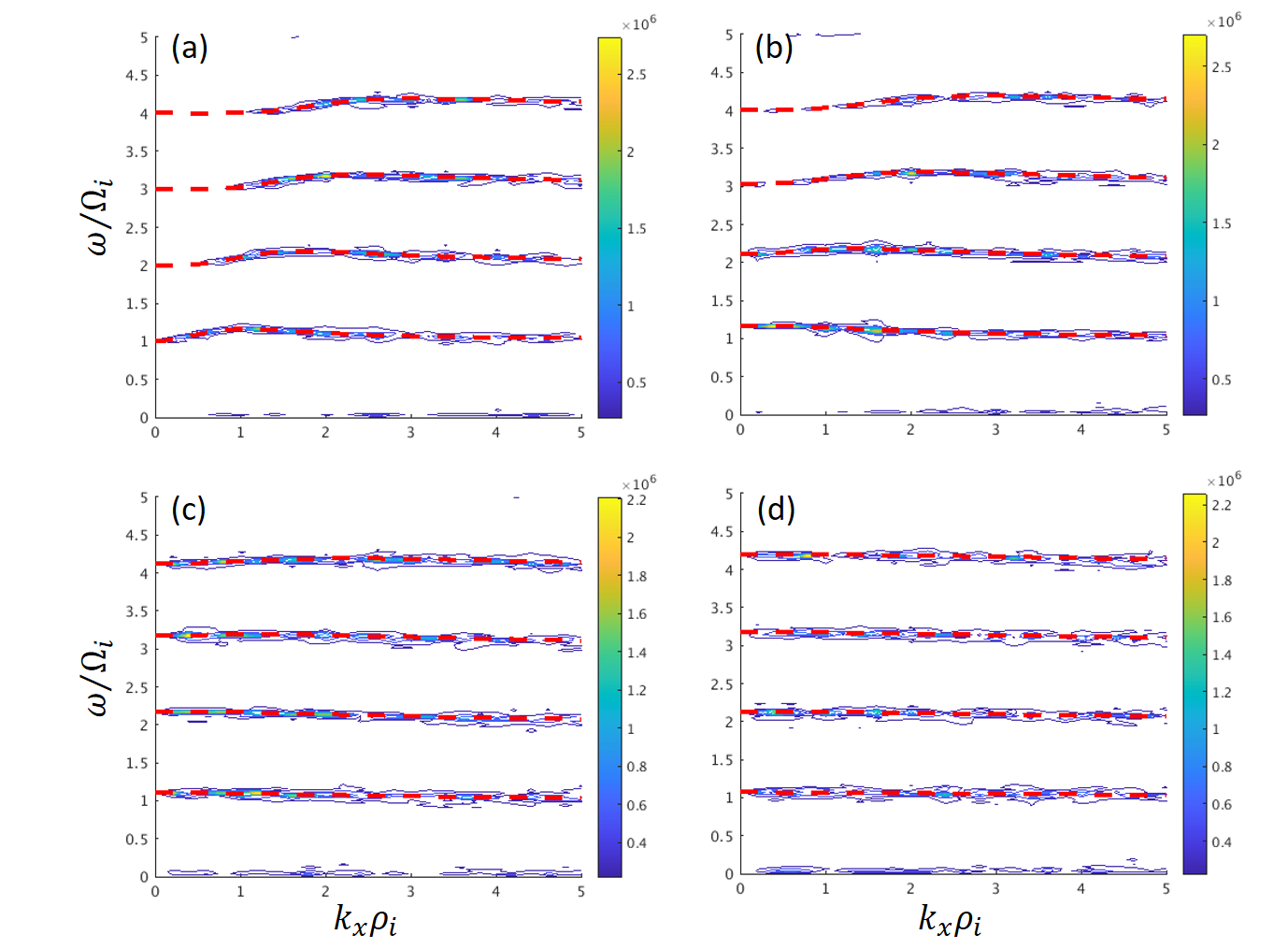}

\caption{\label{fig:wk}Dispersion relation of the IBW in a 2D rectangular
domain. The contour plot of $\tilde{\phi}(k_{x},\omega)$ at (a) $k_{y}\rho_{i}$=0,
(b) $k_{y}$$\rho_{i}$=1.0, (c) $k_{y}$$\rho_{i}$=2.0, and (d)
$k_{y}\rho_{i}$=3.0. The red dashed lines are the theoretical dispersion
relation.}

\end{figure}

During the simulation, the electric potential $\phi$ on each vertex
is recorded. To analyze the data using the Fast-Fourier Transform
(FFT), the $\phi$ on a rectangular mesh is interpolated from the
its values on the vertices of the unstructured mesh. The dispersion
relations of the IBW can be inferred from $\tilde{\phi}$($k_{x}$,$k_{y}$,$\omega$),
where $k_{x}$ and $k_{y}$ are the wave numbers in the $x$- and
$y$-directions, and $\omega$ is the angular frequency. Figure \ref{fig:wk}
plots the contours of the spectral power of $\tilde{\phi}$, and the
contour peaks show the dispersion relation $k_{x}$-$\omega$ at a
fixed $k_{y}$, where $k_{x}$ and $k_{y}$ are normalized to the
ion gyroradius $\rho_{i}$, and $\omega$ to the ion gyrofrequency
$\Omega_{i}$. The contour plots of $\tilde{\phi}$ at $k_{y}\rho_{i}$=0,
1.0, 2.0, 3.0 are shown in Figs.\,\ref{fig:wk}(a)-(d), respectively.
The contour plots are compared with the theoretical dispersion relation
with kinetic ions and adiabatic electrons \citep{Sturdevant2016},
\begin{equation}
1+\theta\sum_{n=-\infty}^{\infty}\frac{n\Omega_{i}\Gamma_{n}(b)}{\omega+n\Omega_{i}}=0.\label{eq:th_dis_re}
\end{equation}
Here, $\theta=q_{i}T_{e}/q_{e}T_{i}$, $b\equiv(k_{\perp}\rho_{i})^{2}$,
$\Gamma_{n}\equiv I_{n}(b)e^{-b},$ $k_{\perp}$ is the perpendicular
wave number, and $I_{n}$ is $n$-th modified Bessel function of the
first kind. For the present 2D simulation, $k_{\perp}=\sqrt{k_{x}^{2}+k_{y}^{2}}$.
The dispersion relation in terms of $(k_{x},k_{y},\omega)$ can be
directly compared with the dispersion relation from the PIC simulation.
The red-dashed lines in Fig.\,\ref{fig:wk} are the dispersion relation
curves at $k_{y}\rho_{i}$=0, 1.0, 2.0, 3.0 from Eq.\,(\ref{eq:th_dis_re}).
The dispersion relation from the PIC simulations agree well with the
theory. 

\subsection{IBW in a 2D circular region with fixed boundary conditions \label{subsec:IBW2}}

\begin{figure}
\includegraphics[scale=0.6]{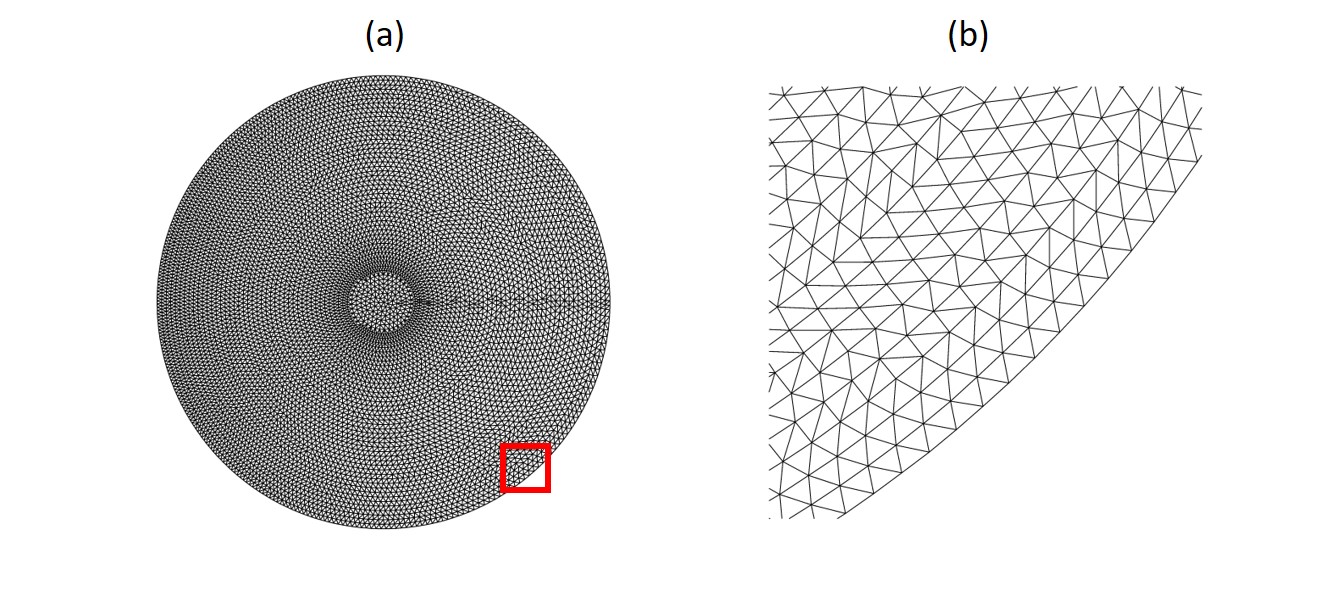}

\caption{\label{fig:cm}(a) The 2D circular simulation domain. (b) The zoom-in
of the red rectangular region in (a).}

\end{figure}

In this section, we simulate the IBW in a 2D circular domain using
an unstructured mesh. The simulation domain is shown in Fig.\,\ref{fig:cm}(a),
and Fig.\,\ref{fig:cm}(b) is the zoom-in of the red rectangular
region in Fig.\,\ref{fig:cm}(a). The simulation domain has 7477
vertices and 14646 triangles. The physical parameters are the same
as in Sec.\,\ref{subsec:IBW1}. The total number of simulation particles
is 4.9$\times10^{10}$ in order to reduce noise and obtain eigenmode
structures, and the timestep is $\Delta t$=0.02$/\Omega_{i}$. The
boundary condition is that $\phi$=0 at the most-outside vertices,
and particles are reflected when entering the most-outside triangles.

\begin{figure}
\includegraphics[scale=0.6]{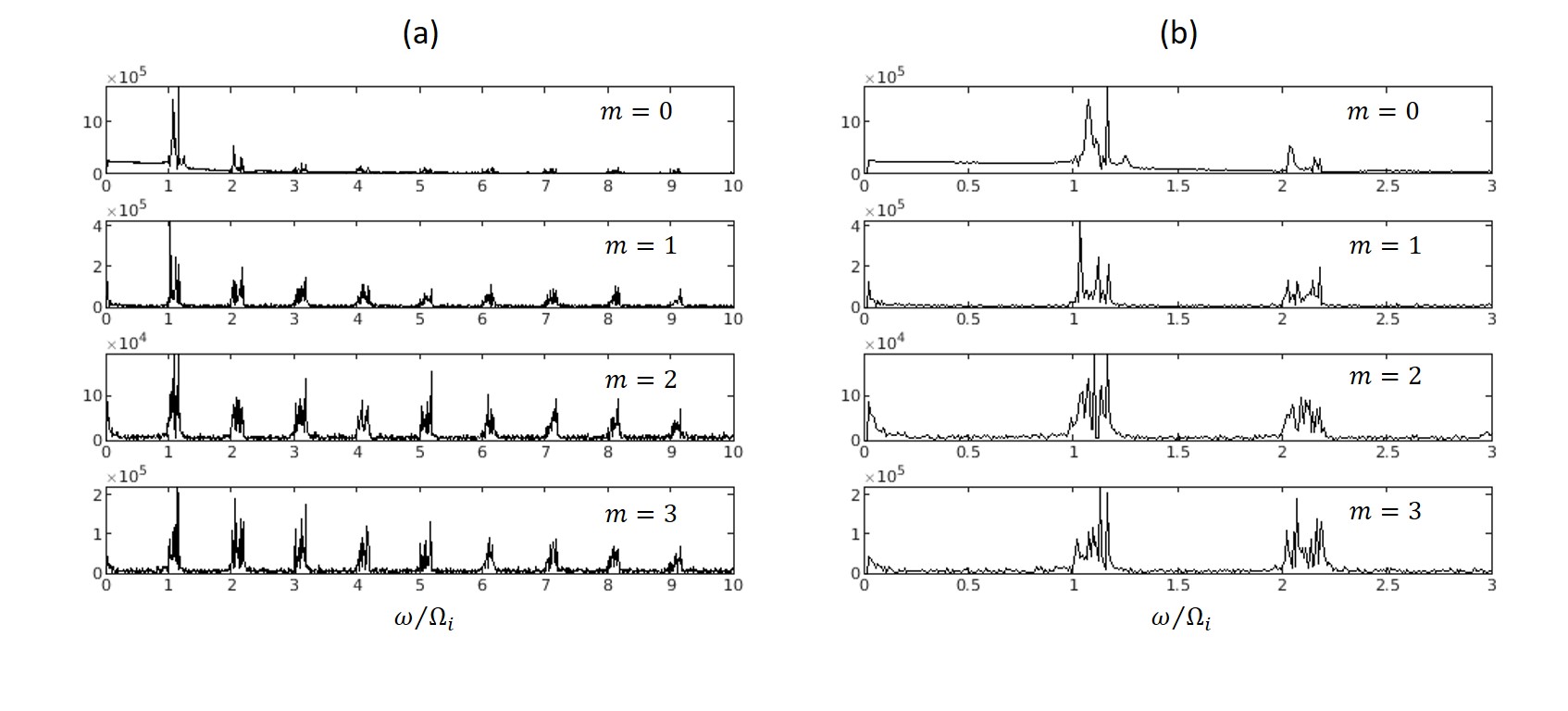}

\caption{\label{fig:cp} (a) Spectrum of the IBW for $m=0,1,2,3$. (b) Zoom-in
of (a) in $0\protect\leq\omega/\Omega_{i}\protect\leq$3. }
\end{figure}

\begin{figure}
\includegraphics[scale=0.6]{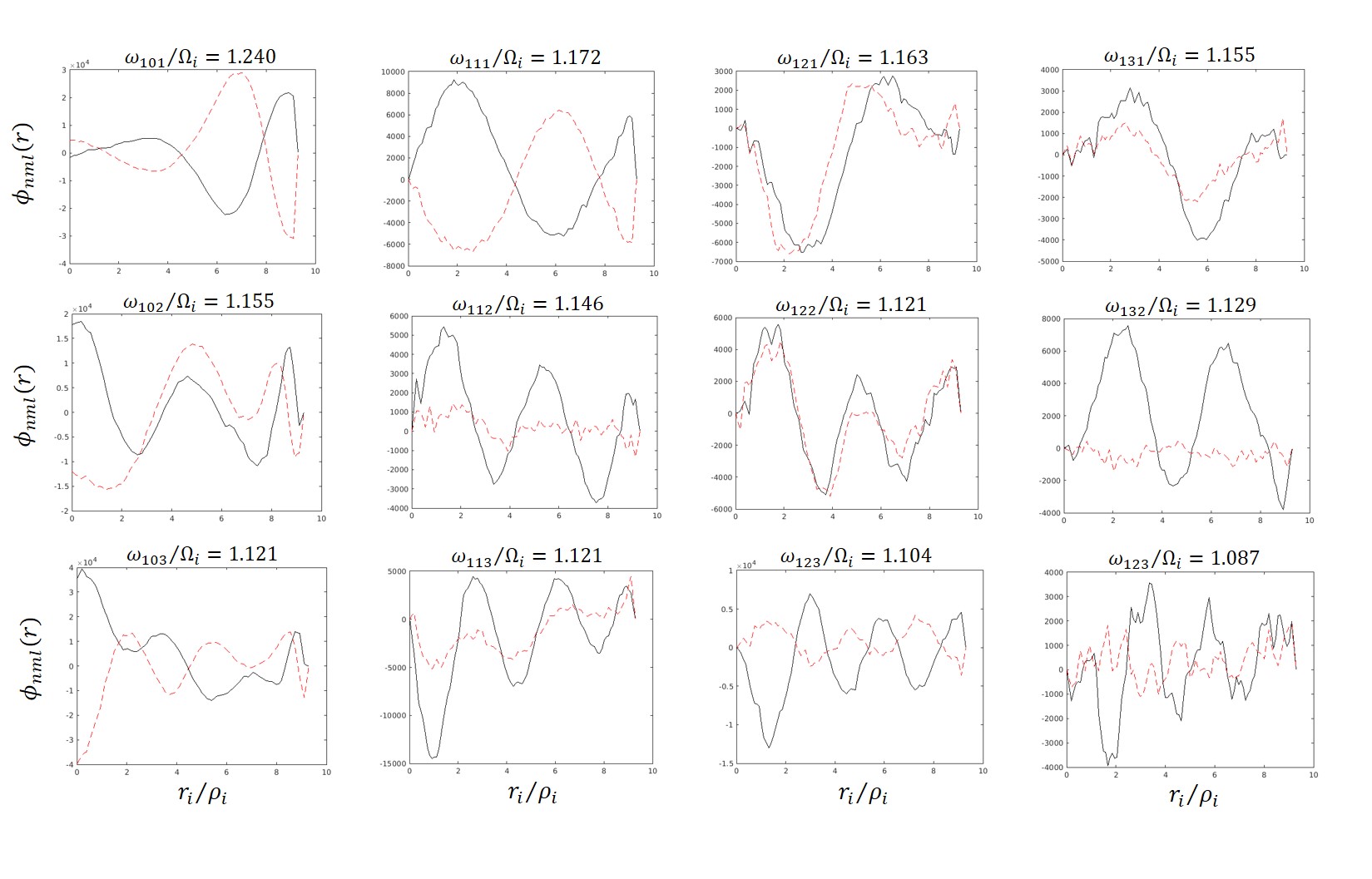}

\caption{\label{fig:eigen}Eigenmode structures $\tilde{\phi}_{nml}(r)$ for
$m=0,1,2,3$ and $l=1,2,3$ in the neighborhood of the first harmonic
($n=1$). Black solid line is real part of $\tilde{\phi}_{nml}(r)$,
and red dashed line is imaginary part. The eigen frequency $\omega_{nml}$
for each eigenmode is listed. }
\end{figure}

For the FFT analysis, $\phi$ is interpolated to a diagnostic circular
mesh which has 101 co-concentric circles with equal intervals, and
61 grid points are distributed on each circle with the same angular
interval. On the diagnostic mesh, $\phi$ is a function of time $t$
and the radial coordinates $(r,\theta)$. Performing the FFT analysis
of $\phi(r,\theta,t)$ along the $\theta$- and $t$-directions, we
discover that for each azimuthal mode number $m$, the spectrum of
the system has a rich structure that can be labeled by two integer
indices, $n$ and $l$. In the neighborhood of each integer harmonics
of $\Omega_{i}$ labeled by $n,$ there exists a family of eigenmodes
labeled by $l.$ The value of $l$ indicates number of oscillations
of the eigenmode in the radial direction. Thus, the eigenmode expansion
for $\phi$ is 

\begin{equation}
\phi=\sum_{n,m,l}\tilde{\phi}_{nml}(r)\textrm{exp}(im\theta-i\omega_{nml}t),
\end{equation}
where $\tilde{\phi}_{nml}(r)$ is the eigenfunction of the mode at
$\omega=\omega_{nml}$. Figure \ref{fig:cp}(a) plots the spectrum
of $\tilde{\phi}$ for $m=0,1,2,3$, where the frequency $\omega$
is normalized to ion gyro-frequency $\Omega_{i}$. Figure \ref{fig:cp}(b)
is zoom-in of the spectrum in the range of $0\leq\omega/\Omega_{i}\leq3$.
The eigenmode structures $\tilde{\phi}_{nml}(r)$ for $m=0,1,2,3$,
$l=1,2,3$ in the neighborhood of the first harmonic ($n=1$) are
shown in Fig.\,\ref{fig:eigen}. We are not aware of any previous
study of these eigenmodes of the IBW in a circular domain. 

\section{Comparison of PIC methods on unstructured meshes}

As discussed in Sec.\,\ref{sec:Introduction}, previous PIC methods
on unstructured meshes used identical shape function for both charge
deposition and field interpolation\citep{Celik2003,Spirkin2004}.
In this section, the energy conservation property of our PIC method
(Method A) is compared with that of a PIC method (Method B) using
identical shape function for both charge deposition and field interpolation.
The comparison is carried out on an unstructured mesh as in Fig.\,\ref{fig:mrec}.

\begin{figure}
\includegraphics[scale=0.5]{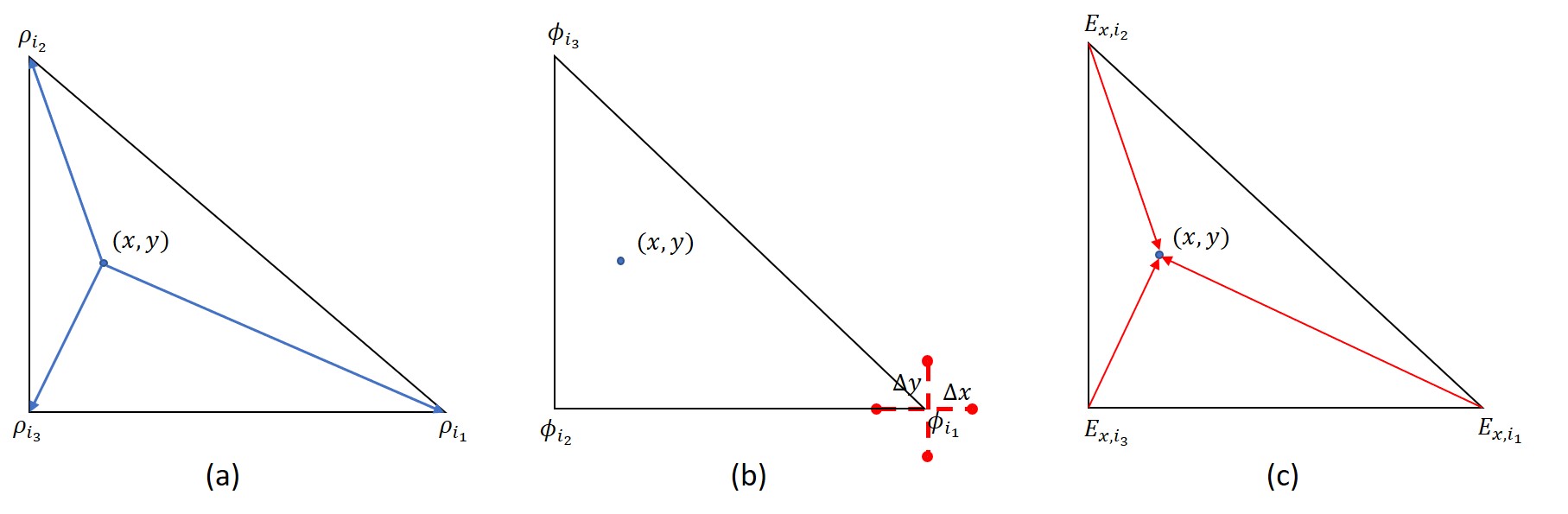}

\caption{\label{fig:fd}A previous PIC method (Method B) on unstructured meshes
with identical shape function for charge deposition and field interpolation.
(a) Depositing the particle's charge into the triangular vertices.
(b) Computing $E_{x}$ and $E_{y}$ by a five-point finite difference
method. (c) Interpolating $\mathbf{E}$ form vertices to the particle's
position.}
\end{figure}

Method B on the unstructured mesh is illustrated in Fig.\,\ref{fig:fd}.
Suppose a particle at $(x,y)$ is inside a triangle. The charge of
the particle is deposited to the triangular vertices by the linear
barycentric functions, as shown in Fig.\,\ref{fig:fd}(a). The electric
potential $\phi_{I}$ is calculated from $\rho_{I}$ by Eq.\,(\ref{eq:disphi}).
With the $\phi_{I}$ on each vertex, the electric field $E_{x}$ and
$E_{y}$ are calculated by a centered finite difference method \citep{Birdsall1991}.
At ($x_{i_{1}}$,$y_{i_{1}}$), the $x$-component of the electric
field is 
\begin{equation}
E_{x}={\displaystyle \frac{\phi(x_{i_{1}}+\Delta x,y_{i_{1}})-\phi(x_{i_{1}}-\Delta x,y_{i_{1}})}{2\Delta x}},
\end{equation}
where $\Delta x$, $\Delta y$ is chosen to be a small value comparing
to the averaged length of triangular edges. To calculate $\phi(x_{i_{1}}\pm\Delta x,y_{i_{1}})$,
we first determine the triangle in which the point $(x_{i_{1}}\pm\Delta x,y_{i_{1}})$
locates, and then interpolating $\phi$ at $(x_{i_{1}}\pm\Delta x,y_{i_{1}})$
from its values on the triangular vertices using linear barycentric
functions, as shown in Fig.\,\ref{fig:fd}(b). $E_{y}$ can be calculated
using a similar method. As Fig.\,\ref{fig:fd}(c) shows, to advance
the particle's position and velocity, the $\mathbf{E}$ at the particle's
position $(x,y)$ is obtained by interpolating the $\mathbf{E}$ at
the triangular vertices to $(x,y)$ using the linear barycentric function.
Note that the Method B uses the linear barycentric functions for both
charge deposition and field interpolation.

Contrast simulations of Methods A and B are carried out on the unstructured
mesh with the periodic boundary conditions. The horizontal length
of the mesh is 10 times larger than the vertical length. The mesh
has 2111 vertices and 4000 triangles. For Method B, $\Delta x$ is
chosen to be 1/10 of averaged length of triangular edges. The physical
parameters are the same as in Sec.\,\ref{subsec:IBW1}. For both
simulations, the timestep is $\Delta t$=0.02$/\Omega_{i}$ and the
number of simulation particles is 1.28$\times10^{6}$. Both simulations
are performed to the time length of 2400$/\Omega_{i}$.

\begin{figure}
\includegraphics[scale=0.5]{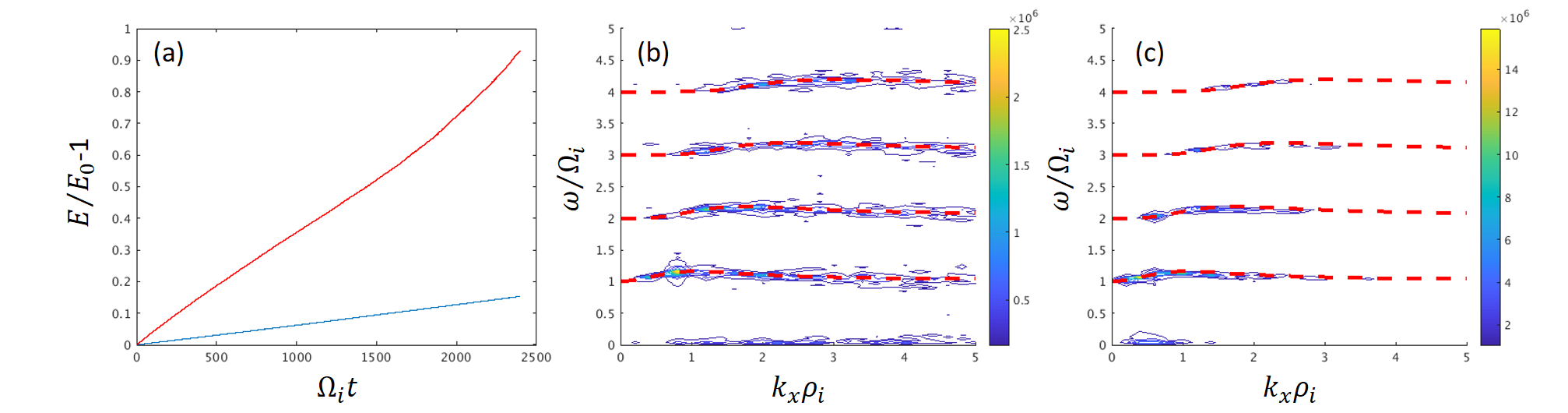}

\caption{\label{fig:compr_err_un}Comparison between our algorithm derived
from the discrete variational principle (Method A) and a previous
method using identical shape function for charge deposition and field
interpolation (Method B). (a) The total energy error during the simulation.
The blue curve is for method A and the red curve is for Method B.
(b) The dispersion relation obtained from Method A, and the red dashed
lines are the theoretical results. (c) The dispersion relation obtained
from Method B.}
\end{figure}

Figure \ref{fig:compr_err_un}(a) shows the comparison of the energy
conservation property of Methods A and B. The energy of the system
consists of the kinetic energy of particles and the potential energy
of the field. We measure the error of total energy during the simulations.
The total energy includes plasma kinetic energy, particles' potential
energy and electric field energy. The blue line is the total energy
error of Method A, and the red line is that of Method B. The growth
rate of total energy error of Method B is 6 times faster than that
of Method A. The IBW dispersion relation for $k_{y}$=0 at $t\sim1800/\Omega_{i}$
is shown in Figures \ref{fig:compr_err_un}(b) and \ref{fig:compr_err_un}(c)
for Method A and Method B, respectively. The dispersion relation from
Method A agrees wells with the theoretical results. In contrast, the
contour plot from Method B cannot recover the dispersion relation
for all harmonics in the regime of $k_{x}\rho_{i}\geq$3. Method A
has a much smaller total energy error, and thus generates a more accurate
dispersion relation. This comparison study demonstrates that our new
algorithm derived from the discrete variational principle has a much
better energy conservation property than previous methods using identical
shape function for charge deposition and field interpolation. We want
to point out that, many PIC methods adopt identical shape function
for both charge deposition and field interpolation in order to conserve
momentum. These PIC methods are commonly referred as ``momentum conserving''
method. However, as Refs.\,\citep{Hockney1981,Birdsall1991} stated,
besides using identical shape function for both charge deposition
and field interpolation, a PIC method must have ``correctly space-centered
difference approximation to derivatives'' \citep{Hockney1981} or
``left-right symmetry'' \citep{Birdsall1991} to have momentum conservation.
Solely adopting identical shape function for both charge deposition
and field interpolation, as in method B and the previous PIC methods
\citep{Celik2003,Spirkin2004,Gatsonis2009,Day2011,Han2016}, can not
guarantee momentum conservation on a unstructured mesh. 

\section{Conclusions and discussion}

In conclusion, we have extended the geometric electrostatic PIC method
in Ref.\,\citep{Xiao2019} to unstructured meshes in 2D simulation
domains. The PIC method uses kinetic ions and adiabatic electrons.
The discrete variational principle gives an algorithm that deposits
particle charge to triangular vertices by Whitney 0-froms and interpolates
the electric field at particles' positions using its values on the
triangular edges and Whitney 1-forms. The formula of Whitney-forms,
the DEC method that computes the discrete electric field on triangular
edges, and the algorithm of charge-deposition and field-interpolation
are described in details. The PIC method has been used to investigate
IBWs on unstructured meshes with two different geometries and boundary
conditions. For the case with periodic boundary conditions in a rectangular
domain, the simulated dispersion relation agrees well with the theory.
For the case on a 2D circular unstructured mesh with fixed boundary
conditions, the spectrum and eigenmode structures are obtained. The
simulation results of our PIC algorithm are compared with those of
a previous PIC method using identical shape function for charge-deposition
and field-interpolation. The comparison shows the new algorithm has
a much better energy conservation property and thus can give a more
accurate dispersion relation. The present paper focuses on demonstrating
the new charge-deposition and field-interpolation methods on unstructured
meshes and has not implemented the symplectic structure-preserving
integration algorithm. The topic of symplectic structure-preserving
integration on unstructured meshes will be addressed in future studies.
\begin{acknowledgments}
This research was supported by the U.S. Department of Energy Office of Science ASCR and FES through SciDAC-4 Partnership Center for High-fidelity Boundary Plasma Simulation (HBPS), under the contract number DE-AC02-09CH11466 through Princeton University. The computing resources were provided on the PPPL computer Traverse operated by the Princeton Institute for Computational Science and Engineering (PICSciE) and on the leadership class computer Cori at NERSC. The data that support the findings of this study are available from the corresponding author upon reasonable request.
\end{acknowledgments}

\bibliographystyle{apsrev4-1}
\bibliography{Refs}

\end{document}